\documentclass[sigconf]{acmart}

\setcopyright{none}

\copyrightyear{2022}
\acmYear{2022}
\setcopyright{usgovmixed}\acmConference[CIKM '22]{Proceedings of the 31st ACM International Conference on Information and Knowledge Management}{October 17--21, 2022}{Atlanta, GA, USA}
\acmBooktitle{Proceedings of the 31st ACM International Conference on Information and Knowledge Management (CIKM '22), October 17--21, 2022, Atlanta, GA, USA}
\acmPrice{15.00}
\acmDOI{10.1145/3511808.3557563}
\acmISBN{978-1-4503-9236-5/22/10}
\usepackage[utf8]{inputenc}
\usepackage{longtable}
\usepackage{threeparttable}
\usepackage{xspace}
\usepackage{appendix}
\usepackage{booktabs}
\usepackage[utf8]{inputenc}
\usepackage{subfig}
\usepackage{textcomp}
\usepackage{amsmath,amsthm}
\usepackage{enumitem}
\usepackage{algorithm}
\usepackage[noend]{algpseudocode}
\algtext*{EndProcedure}%
\usepackage{xcolor,colortbl}
\usepackage{footnote}
\usepackage{url}
\usepackage{tabularx}
\usepackage{multicol}
\usepackage{multirow}
\usepackage{pifont}
\usepackage{bm}
\usepackage{listings}
\usepackage{wrapfig}
\usepackage{relsize}
\usepackage{mathtools}
\usepackage{bbm}
\usepackage{blindtext}
\usepackage{bm}
\usepackage{balance}
\usepackage{mdwlist}
\setlist[enumerate]{leftmargin=*}
\setlist[itemize]{leftmargin=*}

\usepackage{wasysym}
\usepackage{pifont}

\definecolor{green}{rgb}{0,0.7,0.3}

\newcommand{\adj}{\mathbf{A}}
\newcommand{\graph}{G}
\newcommand{\coarsenedGraph}{\tilde{G}}
\newcommand{\vertexSet}{\mathcal{V}}

\newcommand{\matA}{\mathbf{A}}

\newcommand{\matchMat}{\mathbf{S}}
\newcommand{\perm}{\overline{\mathbf{S}}}

\newcommand{\cmark}{\ding{51}}%
\newcommand{\xmark}{\ding{55}}%

\newcommand{\assignMat}{\mathbf{P}}
\newcommand{\numLevels}{\mathbf{L}}
\def\etal{\emph{et al.}}

\newcommand{\method}{\text{CAPER}\xspace}

\newcounter{theorem}

\author{Jing Zhu}
\affiliation{\institution{University of Michigan, Ann Arbor}}
\email{jingzhuu@umich.edu}

\author{Danai Koutra}
\affiliation{\institution{University of Michigan, Ann Arbor}}
\email{dkoutra@umich.edu}

\author{Mark Heimann}
\affiliation{\institution{Lawrence Livermore Natl Laboratory}}
\email{heimann2@llnl.gov}

\begin{document}
\fancyhead{}

\begin{abstract}
Network alignment, or the task of finding corresponding nodes in different networks, is an important problem formulation in many application domains. We propose CAPER, a multilevel alignment framework that \underline{C}oarsens the input graphs, \underline{A}ligns the coarsened graphs, \underline{P}roj\underline{e}cts the alignment solution to finer levels and \underline{R}efines the alignment solution. We show that \method can improve upon many different existing network alignment algorithms by \emph{enforcing alignment consistency across multiple graph resolutions: nodes matched at finer levels should also be matched at coarser levels. } \method also accelerates the use of slower network alignment methods, at the modest cost of linear-time coarsening and refinement steps, by allowing them to be run on smaller coarsened versions of the input graphs.  Experiments show that \method can improve upon diverse network alignment methods by an average of 33$\%$ in accuracy and/or an order of magnitude faster in runtime.  \end{abstract}

\title{CAPER: Coarsen, Align, Project, Refine \\ A General Multilevel Framework for Network Alignment} %
\maketitle

\section{Introduction}
\label{sec:intro}
Graphs or networks are foundational representations for relational structure and their analysis is useful in innumerable scientific and industrial applications. In many diverse tasks, such as recommendation across multiple social networks, protein-protein interaction analysis, and database schema matching~\cite{kazemi2016network}, it is necessary to discover meaningful correspondences between nodes in multiple networks.  This general problem is called network alignment.  

Network alignment methods in general have two main limitations.  First, they may overfit to local structural similarity and fail to preserve higher-order measures of matching consistency~\cite{conealign, refina}.  Second, especially the most accurate methods tend to rely on solving challenging optimization problems with high computational complexity, e.g. quadratic or cubic time in the number of nodes in one of the input graphs \cite{gwl, zhang2019kergm, conealign}.  

We argue that multilevel network analysis is a powerful technique for improving network alignment algorithms on both fronts.  Accordingly, we design the first general multilevel framework to pair with any network alignment method, a four-step framework which we call \method: (1) \underline{C}oarsening a graph into multiple levels of varying coarseness, (2) \underline{A}ligning at the coarsest level, and (3) \underline{P}roj\underline{e}cting back to finer levels, and (4) \underline{R}efining the solution at each level.  We can accelerate the use of slow network alignment algorithms by running them on the smaller coarsened graphs, while refining the solutions at multiple levels of structural resolution encourages greater consistency between the local and global structure of matched nodes.  
Our contributions can be summarized as follows:

\begin{itemize}
    \item \textbf{General-Purpose Framework}: We propose an intuitive multilevel framework (\method) in which any network alignment method can be used.  
    \item \textbf{Design Choices and Empirical Success}: We propose and study specific design choices and parameter settings that work well within \method. We provide code and additional supplementary material at \url{https://github.com/GemsLab/CAPER}. 
    \item \textbf{Study of Accuracy and Runtime Tradeoff}: Through complexity analysis and experiments, we show that \method is able to improve accuracy by 33$\%$ on average across multiple datasets and/or is 10x faster runtime than baselines, depending on the properties of the base methods employed. 
     
\end{itemize}

\begin{figure*}[t!]
    \centering
    \vspace{-0.35cm}
    \includegraphics[width=0.9\textwidth]{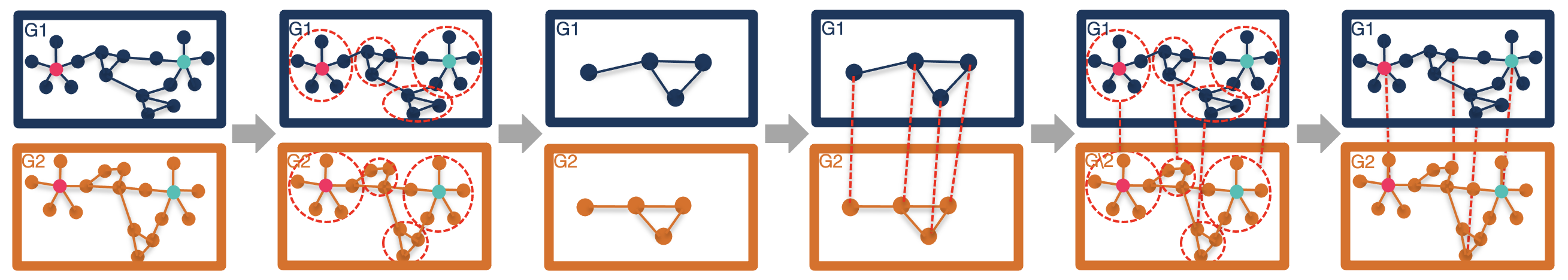}
    \vspace{-0.35cm}
    \caption{An illustrative example of \method.  The pink and blue nodes in the leftmost figure have the same local structural similarity, so methods that overfit the local structural similarity may misalign them. But with the help of higher-order information (coarsened graphs in step 4), \method is able to eventually correctly align the pink nodes as well as the blue nodes.}
    \label{approach}
    \vspace{-0.3cm}
\end{figure*}

\section{Related Work}
\label{sec:related}

\noindent \textbf{Graph Coarsening and Multilevel Methods.}
Graph coarsening~\cite{liu2018graph} is the process of shrinking a large graph into a similar smaller one, such that some properties or structures are preserved, e.g. spectral graph properties or cliques.
It has been used to accelerate many graph mining tasks, including graph clustering~\cite{dhillon2007weighted}, node embedding~\cite{mile,graphzoom} and graph neural networks~\cite{groupinn}. 

\vspace{0.1cm}
\noindent \textbf{Network Alignment.} We focus on unsupervised approaches requiring no known matchings \emph{a priori}.  These can be categorized into two groups. (1)~Classic graph alignment approaches often formulate an \textbf{optimization-based assignment} problem.  
FINAL~\cite{final} optimizes a topological consistency objective which may be augmented with node and edge attribute information.  MAGNA~\cite{magna}, applied to biological networks, uses genetic algorithms to evolve network populations while maximizing proximity consistency criteria. 
More recently, Zhang \etal~\cite{zhang2019kergm} leveraged kernel methods to solve the quadratic assignment problem, but requires cubic computational complexity. (2)~Another line of work relies on \textbf{embedding-based methods.}  
REGAL~\cite{regal} matches structural node embeddings~\cite{jin2021toward, rossi2020proximity} that are directly comparable across networks.  CONE-Align~\cite{conealign} uses embeddings modeling proximity within each graph~\cite{rossi2020proximity} and aligns the graphs' embedding spaces with a subspace alignment procedure,
while GWL~\cite{gwl} solves a Gromov-Wasserstein optimization problem to jointly find node embeddings and the graph matching. G-CREWE~\cite{gcrewe} uses graph compression to accelerate the matching step of embedding-based alignment, though the embedding step is performed on the entire input graphs.  

\begin{table}[t]
\caption{Comparing alignment meta-frameworks.
} 
\label{tab:qualitative}
\centering
\vspace{-0.4cm}
\resizebox{.8\columnwidth}{!}{
\begin{tabular}{lcccc}
\toprule
  \textbf{ } & \rotatebox[origin=c]{0}{\textbf{\bf General }}
  & \rotatebox[origin=c]{0}{\textbf{\bf Multiscale }}
  &  \rotatebox[origin=c]{0}{\textbf{\bf Improves }} & 
  \rotatebox[origin=c]{0}{\textbf{\bf Improves }} \\  
  & & 
  &  \rotatebox[origin=c]{0}{\textbf{\bf  accuracy}} & 
  \rotatebox[origin=c]{0}{\textbf{\bf  runtime}} \\ \hline 
  MOANA~\cite{moana} &  \xmark & \color{green}{\cmark} & \xmark & \color{green}{\cmark} \\
  Boosting~\cite{kyster2021boosting} &  \xmark & \xmark & \color{green}{\cmark} & \xmark \\
  RefiNA~\cite{refina} & \color{green}{\cmark} & \xmark & \color{green}{\cmark} & \xmark \\ 
  \midrule
  \method  & \color{green}{\color{green}{\cmark}} & \color{green}{\cmark} & \color{green}{\cmark} & \color{green}{\cmark} \\
\bottomrule
\end{tabular}
}
\vspace{-0.7cm}
\end{table}

\vspace{0.1cm}
\noindent \textbf{Network Alignment Meta-Frameworks.} A few recent works have proposed meta-frameworks to improve unsupervised network alignment algorithms.  This includes MOANA, the only other multilevel network alignment approach~\cite{moana}.  MOANA uses multiresolution matrix factorization to accelerate FINAL~\cite{final} (it produces negative-valued adjacency matrices that do not work with other network alignment methods) at the cost of some accuracy.   RefiNA~\cite{refina} makes the opposite tradeoff, enforcing greater local consistency to increase accuracy of several base methods at the cost of adding additional runtime. 
Another meta-framework~\cite{kyster2021boosting} studies how design choices of recent embedding-based network alignment methods can be combined to increase accuracy via boosting. Meanwhile, our approach inherits all these benefits, as shown in Tab.~\ref{tab:qualitative}. 
\section{Preliminaries}
\label{sec:preliminaries}

\noindent \textbf{Graphs.} We consider two graphs $\graph_1$ and $\graph_2$
with nodesets $\vertexSet_1,\vertexSet_2$ 
and adjacency matrices $\matA_1, \matA_2$ containing edges between nodes. 
A graph $\graph_i$ has a coarsened version $\coarsenedGraph_i$ with a smaller nodeset $\tilde{\vertexSet}_i$ of $\tilde{n} < n$ nodes.  Each node in the original graph corresponds to a node in the coarsened graph, represented by an assignment matrix $\assignMat \in \{0,1\}^{n \times \tilde{n}}$. For clarity, we drop the $\tilde{\graph}$ notation unless it is necessary to distinguish coarsened and uncoarsened versions.

\vspace{0.1cm}
\noindent \textbf{Alignment.}
An alignment between the nodes of two graphs
can be represented by a matrix $\matchMat$, where $s_{ij}$ is the (real-valued or binary) similarity between node $i$ in $G_1$ and node $j$ in $G_2$. 

\vspace{0.1cm}
\noindent \textbf{Problem Statement.}
Given two graphs $\graph_1$ and $\graph_2$ with meaningful node alignments, but none known \textit{a priori}, 
we seek to shrink them into coarsened versions $\coarsenedGraph_1$ and $\coarsenedGraph_2$, and recover their alignment $\matchMat$ from $\tilde{\matchMat}$ obtained by aligning their coarser versions $\coarsenedGraph_1$ and $\coarsenedGraph_2$. 
\section{Method}
\label{sec:method}
Next, we detail our \method framework, the first general-purpose multilevel framework for unsupervised network alignment that can  accommodate any  base network alignment approach. 
It consists of four steps that are carefully designed in order to achieve higher accuracy and/or lower runtime compared to its base alignment methods: \underline{C}oarsen, \underline{A}lign, \underline{P}roj\underline{e}ct, \underline{R}efine (\method). %
In Fig. \ref{approach}, we provide an example of how \method can implicitly enforce higher-order structural consistency that improves network alignment. 

\subsection{Graph Coarsening}
Given an input graph $\graph_i$, we want to obtain a coarsened graph $\coarsenedGraph_i$ using grouping-based coarsening methods. We leverage the normalized heavy-edge matching (NHEM) heuristic~\cite{dhillon2007weighted} for graph coarsening. This approach repeatedly combines pairs of adjacent nodes into a supernode in decreasing order of degree-normalized edge weight~\cite{mile}, which for edge ($u$,$v$) with weight $w_{uv}$ connecting nodes $u$ and $v$ with degrees $d_u$ and $d_v$ respectively is given by $w_{uv}/\sqrt{d_u d_v}$, until no node is left uncombined or the uncombined nodes do not have uncombined neighbors (isolated nodes).  The resulting coarse graph consists of these supernodes, which share an edge if any of the nodes in one supernode shared an edge in the original graph with any of the nodes in the other supernode.

Graph coarsening turns each input graph $\graph_i$ into a coarsened graph $\coarsenedGraph_i$.  We iteratively repeat this coarsening procedure up to $\numLevels$ times to produce a sequence of coarsened graphs $\coarsenedGraph_i^{(0)}, \ldots, \coarsenedGraph_i^{(L)}$, where the first level is the input graph ($\coarsenedGraph_i^{(0)} = \graph_i$), and the coarsest (smallest) graph is $\coarsenedGraph_i^{(L)}$.   Assignments between nodes at consecutive levels $\ell - 1$ and $\ell$ are contained in a matrix $\assignMat_i^{(\ell)} $ for $\ell \in [1, \ldots, L]$.%

\subsection{Alignment of Coarsened Graphs}

We can apply any  unsupervised network alignment method to align the nodes of the coarsest graphs $\coarsenedGraph_1^{(L)}$ and $\coarsenedGraph_2^{(L)}$ to produce a matching $\matchMat^{(\numLevels)}$.   We observe that the coarsening procedure sometimes generates slightly different numbers of nodes for the same graph even if the input graphs have the same size, so the proposed formulation must be able to handle graphs of different sizes.  This can be done by adding singleton nodes to the smaller graph~\cite{zhang2019kergm, conealign}.  

\subsection{Projection}
We project the alignment solution at the coarsest level $\matchMat^{(\ell)}$ to a mapping between the nodes at the next finer level using the assignment matrices: $\matchMat^{(\ell-1)} = \assignMat_1^{(\ell)^\top} \matchMat^{(\ell)} \assignMat_2^{(\ell)}$.  Note that this solution is coarse, and all nodes in level $\ell - 1$ mapped to the same supernode in level $\ell$ will have the same match.  Thus, we next show how to use the finer graph structure to refine this coarse solution.  

\subsection{Soft Refinement}
\label{sec:refinement}
Recent work for refining network alignment~\cite{refina} operates on ``hard'' initial solutions, where each node is mapped to at most one other node.  Here, we propose a new refinement operator that uses the ``soft'' initial alignments, which better models the various strengths of several potential matches for each node, as shown in Fig. \ref{fig:ablation-soft}.   Given an initial soft (real-valued) alignment $\matchMat$, we iteratively apply the update rule $\matchMat = \text{NORMALIZE} (\matchMat \circ \adj_1 \matchMat \adj_2 + \epsilon)$, where $\circ$ denotes Hadamard product, $\epsilon $ is a small positive minimum matching score to any pair of nodes to prevent over-reliance on the initially discovered matches (we set $\epsilon = 10^{-\lceil \log_{10} \max(n_1, n_2) \rceil}$) and NORMALIZE is  a single round of row-wise then column-wise normalization, as in~\cite{refina}.

\begin{table}[t]
\caption{Dataset statistics: These four datasets represent various phenomena as shown in the description column.}
\label{tab:datasets}
\centering
\vspace{-0.3cm}
\resizebox{1.0\columnwidth}{!}{
\begin{tabular}{l@{\hskip12pt}r@{\hskip12pt}r@{\hskip12pt}r}
\toprule
   \textbf{Name} & \textbf{Nodes} & \textbf{Edges} &  \textbf{Description}  \\
\midrule
     Arenas \cite{koblenz} &  1,133 & 5,451    & communication network \\ 
     
     Hamsterster \cite{koblenz} & 2,426 & 16,613 & social network \\
     
     Facebook \cite{snapnets} & 4,039 & 88,234 & social network \\
     
     Magna \cite{magna} & 1,004 & 8,323 & protein-protein interaction \\
\bottomrule
\vspace{-1.0cm}
\end{tabular}

}
\end{table}

We iteratively apply this project-and-refine procedure between successive levels until we arrive back at the input level, giving us the mapping between nodes in the original graph. 

\subsection{Computational Complexity} 
We analyze the time complexity of \method as a function of the number of nodes $n$ (to simplify notation we assume this is the same for both graphs), for sparse graphs with $O(n)$ edges. Then the 
complexity of our \method framework is $Lf_{\text{coarsen}}(n) + f_{\text{align}}(\frac{n}{2^L}) + L\Big(f_{\text{project}}(n)+ f_{\text{refine}}(n)$\Big). 
The coarsening time applied to each of $L$ levels, $f_{\text{coarsen}}(n)$, is linear in the number of edges using heavy-edge matching~\cite{dhillon2007weighted}, which is $O(n)$.  Projection $f_{\text{project}}$ and refinement $f_{\text{refine}}$ consist of matrix multiplications that, by maintaining a sparse matching matrix, can also run in $O(n)$ time~\cite{refina}. 

Meanwhile, with 
NHEM shrinking the graph by approximately a factor of 2 at each level~\cite{dhillon2007weighted}, note that we are able to run the base alignment step on a smaller graph, incurring a runtime of $f_{\text{align}}(\frac{n}{2^L})$ as opposed to $f_{\text{align}}(n)$ by applying the base alignment algorithm to the full input graphs. Thus, \method can offer computational speedup particularly for slow base alignment methods, where $f_{\text{align}}$ may be asymptotically large (such as $O(n^3)$),%
and the savings may outweigh the overhead of coarsening, projection, and refinement.

\section{Experiments}
\label{sec:experiments}

We first describe our experimental setup and the datasets and baseline methods used in our empirical analysis, and then show quantitative improvements from \method and a closer ablation study.

\label{subsec:exp-setup}

\enlargethispage{\baselineskip}

\vspace{0.1cm}
\noindent \textbf{Data.} 
We use simulated and real alignment scenarios on graphs representing various real-world phenomena (Tab. ~\ref{tab:datasets}). Following prior works~\cite{regal,final,kyster2021boosting}, we simulate a network alignment scenario with known ground truth: a graph with adjacency matrix $\adj$ is aligned to a noisy permuted copy $\adj^* = \perm \adj \perm^\top$ and $\perm$, for which we generate a random permutation matrix $\perm$; we then randomly remove edges from $\adj^*$ with probability $p \in [0.05, 0.10, 0.15, 0.20, 0.25]$. 
The MAGNA~\cite{magna} networks are protein-protein interaction (PPI) networks that are aligned to versions of themselves with various percentages of low-confidence PPIs (edges) added; thus, all edges in this graph represent real-world phenomena and we do not need to synthesize an alignment scenario.

\vspace{0.1cm}
\noindent \textbf{Baselines.} 
\label{sec:baselineset} We use \textbf{(1)} FINAL~\cite{final} and \textbf{(2)} REGAL~\cite{regal}, which are popular unsupervised network alignment methods that have usable public codebases and represent different classes of techniques (optimization and node embeddings), demonstrating the wide applicability of our framework.  We also use a more recent approach, \textbf{(3)}~GWL~\cite{gwl}, which combines optimization and node embeddings, and achieves good accuracy but has slow runtime due to its $O(n^3)$ computational complexity. Moreover, we consider the post hoc refinement method RefiNA applied to each of the network alignment methods: \textbf{(4)}~FINAL-RefiNA, \textbf{(5)}~REGAL-RefiNA and \textbf{(6)}~GWL-RefiNA. Additionally, we  use  \textbf{(7)} MOANA~\cite{moana} as a baseline, the only other multilevel network alignment method.

\begin{figure*}[t!]
	\centering
	
    \subfloat{%
      \includegraphics[width=.9\textwidth]{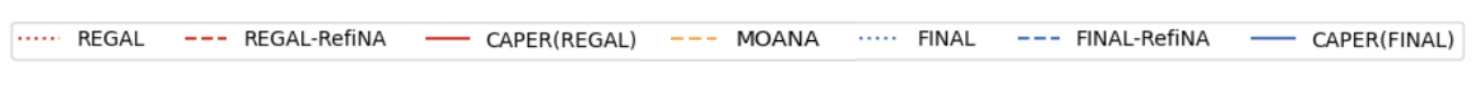}
    }
	\vspace{-0.6cm}	
	\setcounter{subfigure}{0}
	
    \subfloat[Arenas\label{arenas}]{%
      \includegraphics[width=0.23\textwidth,trim=0 0 0 0.85cm, clip]{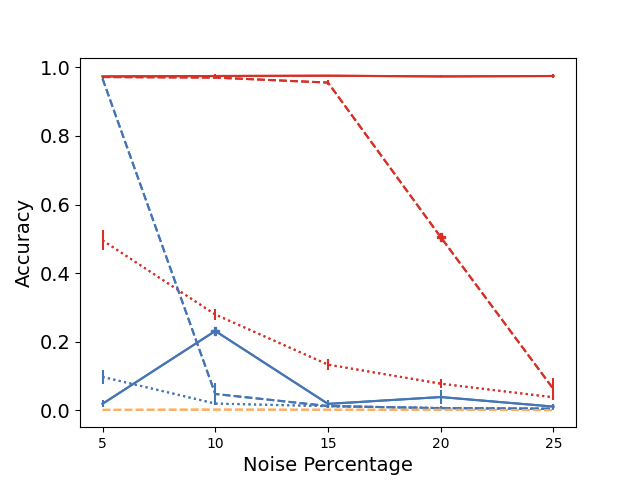}
    }
    ~
    \subfloat[Hamsterster\label{hamster}]{%
      \includegraphics[width=0.23\textwidth,trim=0 0 0 0.35cm, clip]{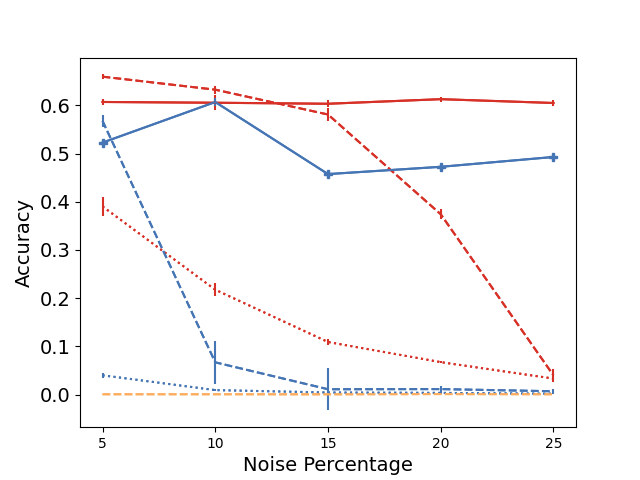}
    }
    ~
    \subfloat[Magna\label{Magna}]{%
      \includegraphics[width=0.23\textwidth,trim=0 0 0 0.35cm, clip]{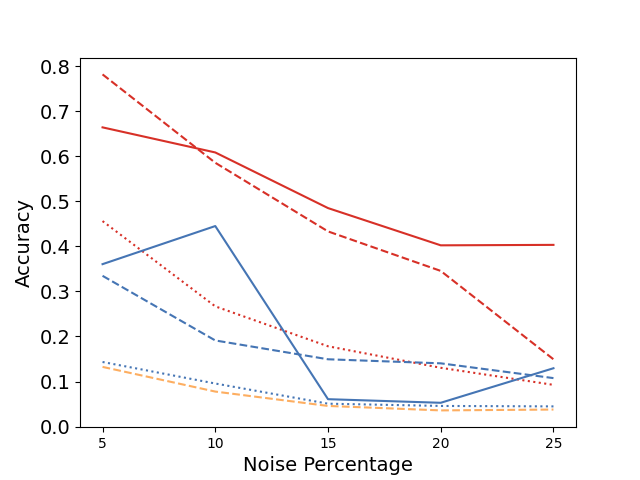}
    }
    ~
    \subfloat[Facebook
    \label{fb}]{\includegraphics[width=0.23\textwidth,trim=0 0 0 0.35cm, clip]{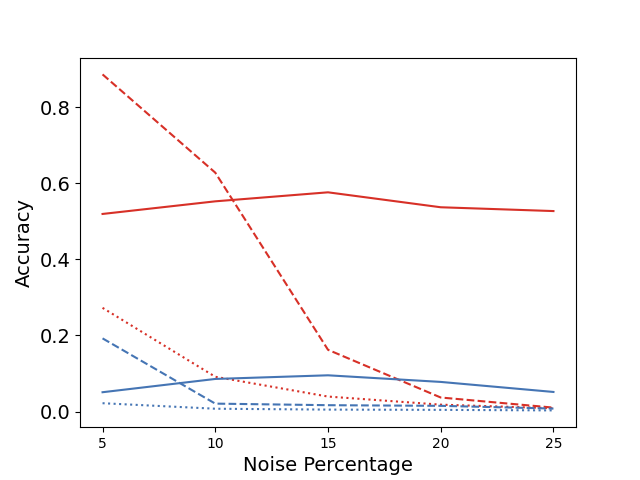}
    }
	\centering
	\vspace{-0.4cm}
\caption{Accuracy (solid lines) vs.\ different noise levels.  \method outperforms baselines, particularly as noise increases.}
    \label{fig:simulated-acc}
\end{figure*}

\begin{figure*}[t!]
\vspace{-0.45cm}
    \centering
    \includegraphics[width=0.92\textwidth]{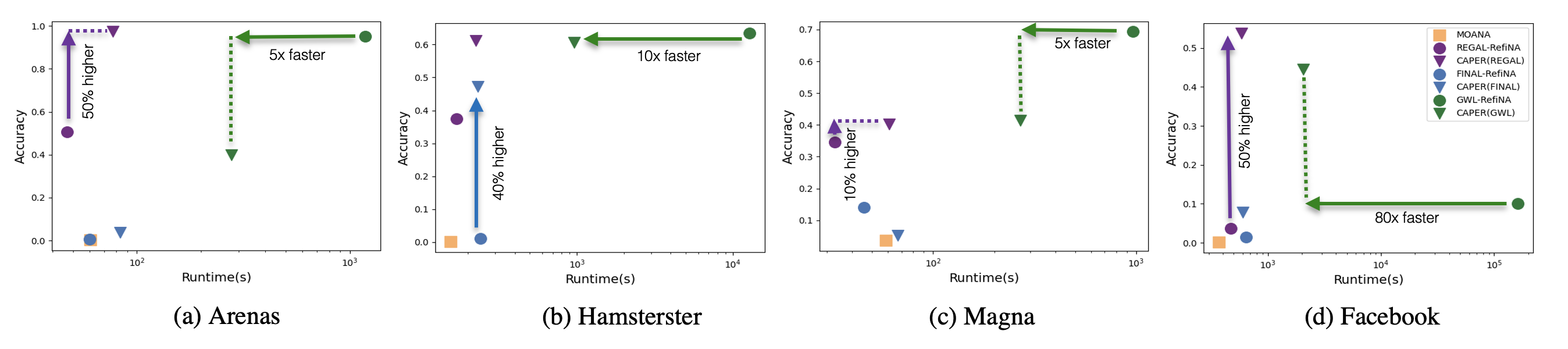}
    \vspace{-0.65cm}
    \caption{Accuracy vs. runtime for \method and RefiNA for 20\% noise. \method yields better accuracy for FINAL and REGAL by enforcing higher-order consistency. For GWL, \method runs up to 80x faster because the alignment is run on smaller graphs.}%
    \label{fig:accrun}
    \vspace{-0.3cm}
\end{figure*}

For FINAL's prior alignment information, we take the top $k=\lfloor \log_2 n\rfloor$ most similar nodes by degree for each node~\cite{regal,conealign}.  We set other parameters for REGAL~\cite{regal} and GWL~\cite{gwl} using the defaults recommended by the authors.  

\vspace{0.1cm}
\noindent \textbf{\method variants.} We test variants of \method using each base alignment method: CAPER(FINAL), CAPER(REGAL), and CAPER(GWL). 
We use 3 coarsening levels and 100 refinement iterations, as in~\cite{refina}, to balance accuracy and computational efficiency (we found that more refinement may increase performance if that is desired and increased runtime is acceptable). 

\noindent \textbf{Evaluation.} We measure \textbf{alignment accuracy}, or the proportion of correctly aligned nodes, and \textbf{runtime}.

\begin{table}[t]
\caption{Number of nodes in the coarsened graph after 2--4 levels of coarsening.}
\label{tab:num-level-ablation}
\vspace{-0.3cm}
\begin{tabular}{l@{\hskip30pt}r@{\hskip20pt}r@{\hskip20pt}r}
\toprule
   \textbf{Name} & \textbf{2} & \textbf{3} &  \textbf{4}  \\
\midrule
     Hamsterster & 1,288 & 702 & 418 \\
     Facebook & 2,078 & 1,078 & 572 \\
\bottomrule
\end{tabular}

\vspace{-0.5cm}
\end{table}

\subsection{Alignment Accuracy}
\label{sec:acc}

\noindent \textbf{Setup.} In Fig.~\ref{fig:simulated-acc}, we report the average accuracy and standard deviation (+ sign: standard deviation $> 0.05$) over five trials for each setting, except for Magna where we do not simulate alignments.

\noindent \textbf{Results}.  While the existing multilevel alignment method, MOANA, has accuracy below its single-level counterpart FINAL as expected, our multilevel framework, \method, significantly outperforms different base alignment methods as well as their single-level refined variants using RefiNA.  Moreover, we can see that \textbf{\method is more robust to noise} due to the multilevel consistency that it encourages; this is especially notable for CAPER(REGAL) whose performance is very stable even when the noise level increases.

\subsection{Alignment Runtime}

\noindent \textbf{Setup \& Evaluation.}
Due to GWL's slow runtime, we only run it for one trial on the largest Facebook dataset. Others are averaged over five trials in Fig.~\ref{fig:accrun}. 

\vspace{0.1cm}
\noindent \textbf{Results.}
For faster base methods such as FINAL and especially REGAL, our improvements are mainly in accuracy (up to 50\% higher accuracy); the computational savings of performing the alignment on smaller graphs does not outweigh the overhead of coarsening and refinement when the base alignment method is fast. However, when the base alignment method is slow, as is the case for GWL, our framework results in considerable computational savings (5-80$\times$ faster), while largely preserving accuracy.  Compared with MOANA, CAPER achieves
better accuracy.

\subsection{Sensitivity Analysis}

\begin{figure}[t]
    \vspace{-0.65cm}
	\setcounter{subfigure}{0}
    ~
    \subfloat[Hamsterster\label{level-hamster}]{%
      \includegraphics[width=0.23\textwidth]{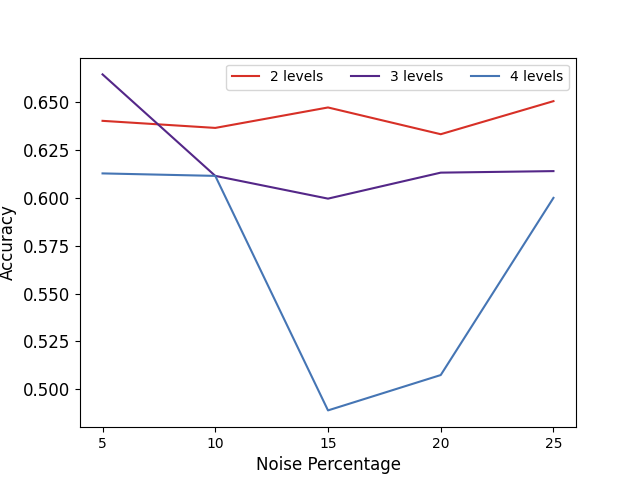}
    }
    ~
    \subfloat[Facebook
    \label{level-fb}]{\includegraphics[width=0.23\textwidth]{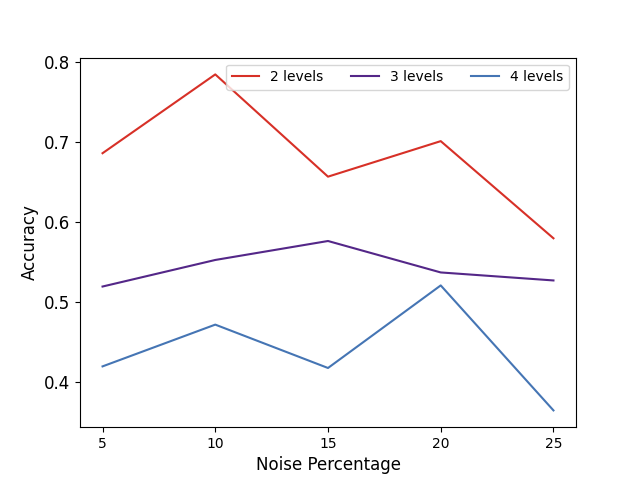}
    }
	\centering
		\vspace{-0.3cm}
\caption{Sensitivity to number of coarsening levels for CAPER(REGAL).
In general, 2 levels leads to highest accuracy. We use 3 levels for the best accuracy/runtime tradeoff. }
    \label{fig:ablation-levels}
    \vspace{-0.45cm}
\end{figure}

\noindent \textbf{Number of levels.} 
Figure~\ref{fig:ablation-levels} compares the performance of CAPER(REGAL) with different numbers of coarsening levels on the Hamsterster and Facebook datasets. %
The number of coarsening levels leads to a tradeoff between accuracy and runtime: more coarsening leads to smaller coarsened graphs (Tab. ~\ref{tab:num-level-ablation}) and faster runtime at a cost of some accuracy.   For our main experiments, we used 3 levels of coarsening for all datasets to balance this tradeoff, and could use 2 levels to achieve even higher accuracy. 

\begin{figure}[t]
	\vspace{-0.59cm}
	\setcounter{subfigure}{0}

    ~
    \subfloat[Magna\label{soft-magna}]{
      \includegraphics[width=0.23\textwidth]{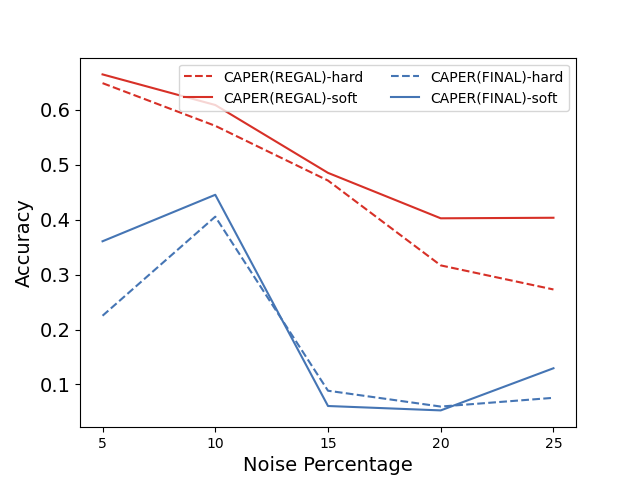}
    }
    ~
    \subfloat[Facebook
    \label{soft-fb}]{\includegraphics[width=0.23\textwidth]{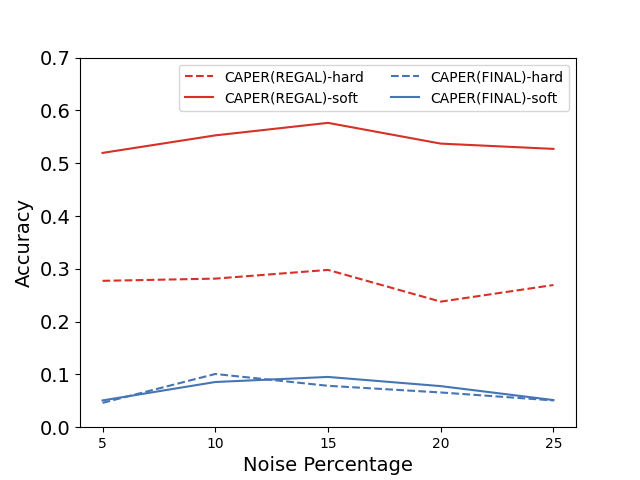}
    }
	\centering
	\vspace{-0.3cm}
\caption{Sensitivity to soft/hard refinement for CAPER(REGAL) and CAPER(FINAL). 
Soft refinement works significantly better, especially for accurate base methods. }
    \label{fig:ablation-soft}
    \vspace{-0.45cm}
\end{figure}

\vspace{0.1cm}
\noindent \textbf{Hard vs. soft refinement.} 
In Fig.~ \ref{fig:ablation-soft}, we see improvement from our refinement of more expressive ``soft'' alignments (\S~\ref{sec:refinement}), most noticeably for the base method REGAL. 
For FINAL, because its initial solution is less accurate on these datasets, we used hard refinement when operating directly on its solution (at the coarsest level) and soft refinement at subsequent levels.  This also explains the smaller gap in performance. 

\section{Conclusion}
\label{sec:conclusion}

We describe the first general-purpose multilevel framework for unsupervised network alignment. It works with various base network alignment algorithms, making them more accurate and robust by incorporating multiscale graph information, and accelerating the runtime by allowing them to operate on smaller input graphs. 
However, not all coarsening methods work well. Some recent spectral coarsening methods \cite{graphzoom} will give clusters with zero nodes and thus our multi-level alignment framework could fail. One possible future direction is to characterize the effect of various coarsening methods on multilevel network alignment.

{\small
\section*{Acknowledgements}
 
This work was performed under the auspices of the U.S. Department of Energy by Lawrence
Livermore National Laboratory under Contract DE-AC52-07NA27344, Lawrence Livermore National Security, LLC. and was supported by the LLNL-LDRD Program under Project No. 21-ERD-012. Jing Zhu was an intern at Lawrence Livermore National Laboratory while working on this project. {It was also partially supported by} the NSF under Grant No. IIS 1845491, and Amazon and Facebook faculty awards. 
} 
\clearpage
\balance
\bibliographystyle{plain}
\bibliography{BIB/bibliography}
\end{document}